\documentclass[prl,twocolumn,groupedaddress,showpacs]{revtex4}
\usepackage{amssymb}
\usepackage{graphicx}
\usepackage[dvips]{epsfig}
\usepackage{bm}
\usepackage{amsmath}
\usepackage{wrapfig}

\usepackage{float}

\usepackage{color}

\begin{document}
\title{Trapped Atoms in One-Dimensional Photonic Crystals}

\author{C.-L. Hung$^\ast$,$^{1,3}$ S. M. Meenehan$^\ast$,$^{2,3}$ D. E. Chang,$^{4}$ O. Painter,$^{2,3}$ and H. J. Kimble$^{1,3}$}
\address{$^1$ Norman Bridge Laboratory of Physics 12-33}
\address{$^2$ Thomas J. Watson, Sr., Laboratory of Applied Physics 128-95}
\address{$^3$ Institute for Quantum Information and Matter, California Institute of Technology, Pasadena, CA 91125, USA}
\address{$^4$ ICFO - Institut de Ciencies Fotoniques, Mediterranean Technology Park, 08860 Castelldefels (Barcelona), Spain}

\date{\today}

\begin{abstract}
We describe one-dimensional photonic crystals that support a guided mode suitable for atom trapping within a unit cell, as well as a second probe mode with strong atom-photon interactions. A new hybrid trap is analyzed that combines optical and Casimir-Polder forces to form stable traps for neutral atoms in dielectric nanostructures. By suitable design of the band structure, the atomic spontaneous emission rate into the probe mode can exceed the rate into all other modes by more than tenfold. The unprecedented single-atom reflectivity $r_0 \gtrsim 0.9$ for the guided probe field should enable diverse investigations of photon-mediated interactions for $1D$ atom chains and cavity QED.
\end{abstract}

\maketitle

New opportunities in Atomic, Molecular, and Optical Physics and Quantum Information Science emerge from the capability to achieve strong radiative interactions between single atoms and the fields of nanoscopic optical waveguides and resonators \cite{Kimble:2008}. For example, strong atom-photon interactions in lithographic structures \cite{vahala-review, Lev:2004, Dayan:2008, hinds, reichel} could be used to create quantum optical circuits with long-range atom-atom interactions mediated by single photons \cite{cirac97,duan04}.
Moreover, linear arrays of atoms radiatively coupled to nanophotonic waveguides exhibit a wide range of remarkable phenomena, including coherent transport of atomic emission \cite{fan05,kien05,dzsotjan10}, guided superradiance and polaritons \cite{Kien2008,Chang2007b,Zoubi2010}, as well as highly reflecting atomic mirrors \cite{changy11,Chang2012}. The interplay of atomic emission into the waveguide and photon-mediated forces can lead to self-organization of atoms into exotic spatial configurations along the waveguide \cite{asboth_optomechanical_2008,Chang2012b}.

A long-standing obstacle to this scientific frontier is the challenge of trapping atoms in vacuum near dielectric surfaces ($\sim 100$nm) while at the same time achieving strong interactions between one atom and photon. A far-off resonance dipole-force trap (FORT) \cite{ye08} can provide atomic localization by using modes of the dielectric for optical trapping \cite{balykin91,vernooy97,burke_designing_2002} and has been used to trap cold atoms within hollow-core optical fibers \cite{Renn1995,Ito1996,Christensen2008,Bajcsy2009} and external to fiber-taper waveguides \cite{Vetsch2010,Dawkins2011,Goban2012}.

\begin{figure}[t!]
\centering
\includegraphics[width=1.0\columnwidth]{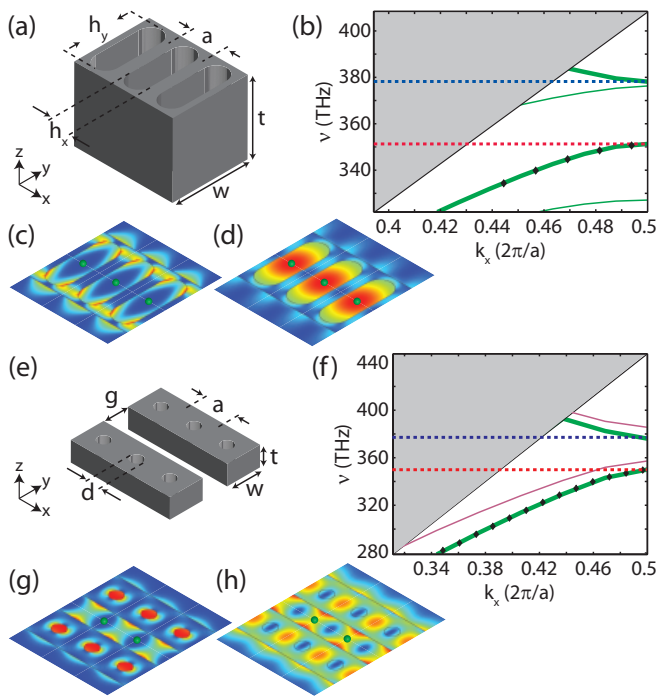}
\caption{\textbf{a)} Schematic for the single nanobeam structure with dimensions $(a,w,t,h_x,h_y) = (367,845,825,246,745)$nm. b) Band diagram for the single nanobeam in a) showing only bands with even vector symmetry about the $y$ and $z$ symmetry planes. The trapping and probing bands are shown as thicker lines, with the trap $\omega_T/2\pi$ (probe $\omega_A/2\pi$) frequency as a blue (red) dashed line. c) Field intensity of the blue trapping mode and d) field amplitude of the probe mode in the center plane $z = 0$ for the single nanobeam in a). Green spheres mark the locations of minima of the trapping potential. \textbf{e)} Schematic for the double nanobeam structure with $(a,w,t,d,g) = (335,335,200,116,250)$nm. f) Band diagram for the double nanobeam in e) displaying only modes of even vector symmetry in $z$. The proximity of the two nanobeams results in a band structure composed of even (green) and odd (magenta) superpositions of single nanobeam modes. We focus on the even parity supermodes due to their large field amplitude in the gap. g) Field intensity of the blue trapping mode and h) field amplitude of the probe mode in the center plane $z=0$ of the double nanobeam in e). The black diamonds in b) [f)] mark resonances for finite structures of $81$ unit cells from Figs. \ref{fig4} [\ref{fig5}].}
\vspace{-4mm}
\label{fig1}%
\end{figure}

Motivated by these advances, in this manuscript we present principles for the design of optical traps and strong atom-photon interactions in one-dimensional ($1D$) photonic crystal waveguides. We analyze the potential $U_{\text{tot}}(\mathbf{r})$ due to light-shifts from a FORT \cite{Rosenbusch2009, Fam2012,Ding2012} together with Casimir-Polder (CP) interactions with the dielectric \cite{Agarwal1975, Hinds1997, Buhmann2004, Rodriguez2009} (Figs. \ref{fig1}(a, e)). Despite the proximity of the surfaces, stable potentials $U_{\text{tot}}(\mathbf{r})$ are achieved for modest optical intensities ($\sim 5$mW/$\mu$m$^2$) for blue-detuned FORTs operated at a `magic' wavelength for the D2 line of atomic Cesium \cite{ye08}. A new possibility for trapping is also identified for which vacuum forces from CP interactions are exploited to close the trap perpendicular to the plane of structure, which would otherwise be unstable with either the FORT or the CP potential alone \cite{Kardar}.

In addition to the waveguide trapping properties, strong near-resonance atom-photon interactions of trapped atoms are found to arise for waveguides with properly tailored band structure \cite{Soljacic2002,Koenderink2006,Rao2007,Baba2008,Hoang2012}. For practically realizable structures, we find $\gamma_{1D}/\gamma^{\prime} \gtrsim 10$, where $\gamma_{1D}$ is the atomic decay rate into the (guided) probe mode and $\gamma^{\prime}$ the rate into all other modes. One atom trapped within the structure could thereby attenuate a resonant probe with transmission $|1-r_0|^2 \lesssim 10^{-2}$ \cite{fan05,Chang2012}.

As illustrated in Fig. \ref{fig1}, we focus on two of the simplest quasi-1D photonic crystal geometries.
The first waveguide consists of a single silicon-nitride nanobeam (refractive index $n=2$) with a $1D$ array of filleted rectangular holes along the propagation direction; atoms are trapped in the centers of the holes (Fig. \ref{fig1}(a)). The second waveguide consists of two parallel silicon nitride nanobeams, each with a periodic array of circular holes, with atoms trapped in the gap between the beams (Fig. \ref{fig1}(e)).

The design of a $1D$-photonic crystal waveguide with distinct modes for optical trapping and strong atom-photon interactions is constrained by the region of the optical band structure containing a continuum of unguided optical modes (i.e., the \textit{light cone} indicated in gray in Fig. \ref{fig1}(b, f)). Modes within the light cone can still have large amplitude in the structure but radiate energy into the surrounding vacuum leading to unacceptable loss. The top of the vacuum light line is at the Brillouin zone boundary ($X$-point, where $k_x a = \pi$), so the lattice constant $a$ is constrained by $a<\lambda/2$, where $\lambda$ is the smaller of the (vacuum) wavelengths for trapping and probe fields.
Here, $k_x$ is the Bloch wavevector along the waveguide axis $x$.

Once $a$ is fixed, additional guided modes can be `pulled' below the light line by increasing the width and thickness of the structure. With appropriate modes below the light line for probing and trapping, the spacing of these modes at the $X$-point can be tuned by altering the size of the holes, which enables the probe mode to be resonant with the frequency $\omega_A$ of the atomic transition while simultaneously matching the optical frequency $\omega_T$ of the trap mode to a `magic' frequency for the atom \cite{ye08}.

Within this general context, here we consider only blue-detuned FORTs for which the trapping mode has an intensity minimum at the trapping site \cite{redFORT}. Our analysis is for the $D2$ line of atomic Cs with probe wavelength near the atomic resonance $\lambda_{A} = 852$nm and with a blue-detuned FORT at the magic wavelength $\lambda_{T} = 793$nm \cite{ye08}. Note that our results are readily transcribed to other atomic transitions by way of the scale invariance of Maxwell's equations \cite{JoannopolousBook}.

The photonic crystals are assumed to be suspended in vacuum and composed of $SiN$. Band structures are calculated using the MIT photonic bands software package \cite{MPB}. Field profiles for guided modes are calculated using the finite-element-method (FEM) simulations \cite{COMSOL}. Results for the single and double nanobeam structures are presented in Fig. \ref{fig1}.

With suitable guided modes for trapping in hand, we have developed numerical tools for evaluating the FORT and CP potentials inside the waveguide, and hence the total potential $U_{\mathrm{tot}}=U_{\mathrm{FORT}}+U_{\mathrm{CP}}$. The adiabatic potential $U_{\mathrm{FORT}}(\mathbf{r})$ is readily calculated using the electric field distribution of the trap mode, $\mathbf{E} (\mathbf{r}) = \mathbf{u}_{k_x} (\mathbf{r}) e^{i k_x x}$ \cite{Rosenbusch2009, Fam2012,Ding2012}. Here $\mathbf{u}_{k_x} (\mathbf{r})$ is the periodic Bloch wave function at propagation constant $k_x$.

The surface potential $U_{\mathrm{CP}}(\mathbf{r}_a)$ is determined from the formalism in Ref. \cite{Buhmann2004} for the imaginary component of the scattering Green's tensor $\mathbf{G}_{sc}(\mathbf{r}_a,\mathbf{r}_a,\omega)$, which is the Green's tensor from Maxwell's equations for a point dipole at the atomic location $\mathbf{r}_a$ with the vacuum contribution (i.e., no dielectric structure) subtracted. We evaluate $\mathbf{G}_{sc}(\mathbf{r}_a,\mathbf{r}_a,\omega)$ numerically by adapting the procedures from Ref. \cite{Rodriguez2009}, as described in \cite{SM}.

\begin{figure}[t!]
\centering
\includegraphics[width=1.\columnwidth]{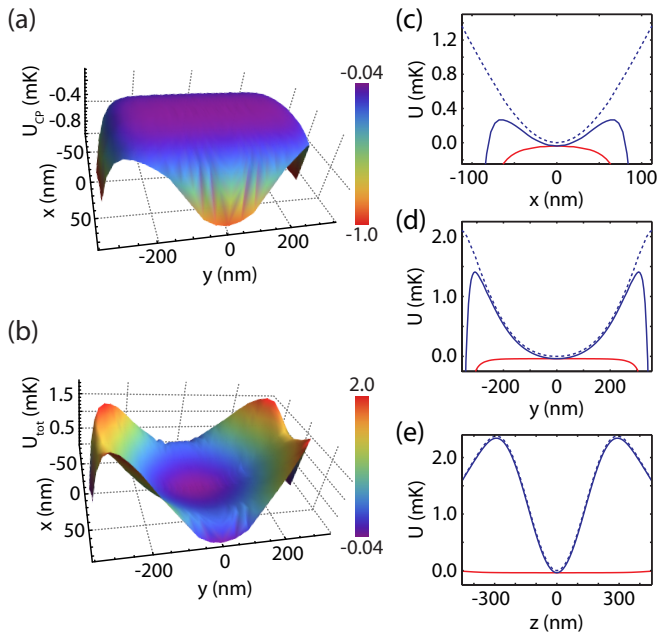}
\caption{Trapping potentials for the single nanobeam structure in Fig. \ref{fig1}(a) for Cs $6S_{1/2}, F = 4$ level and $\lambda_{T} = 793$nm. (a) Casimir-Polder potential $U_{\mathrm{CP}}(\mathbf{r})$ and (b) total potential $U_{\mathrm{tot}}(\mathbf{r})=U_{\mathrm{CP}}(\mathbf{r})+U_{\mathrm{FORT}}(\mathbf{r})$ in the central $z=0$ plane. (c-e) show line cuts of $U_{\mathrm{CP}}$ (red solid), $U_{\mathrm{FORT}}$ (blue dashed), and $U_{\mathrm{tot}}$ (blue solid) along the (c) $x$-, (d) $y$-, and (e) $z$-axis. Average trap intensity for a unit cell is $4.9$mW/$\mu$m$^2$.}
\vspace{-4mm}
\label{fig2}
\end{figure}

\begin{figure}[t!]
\centering
\includegraphics[width=1.0\columnwidth]{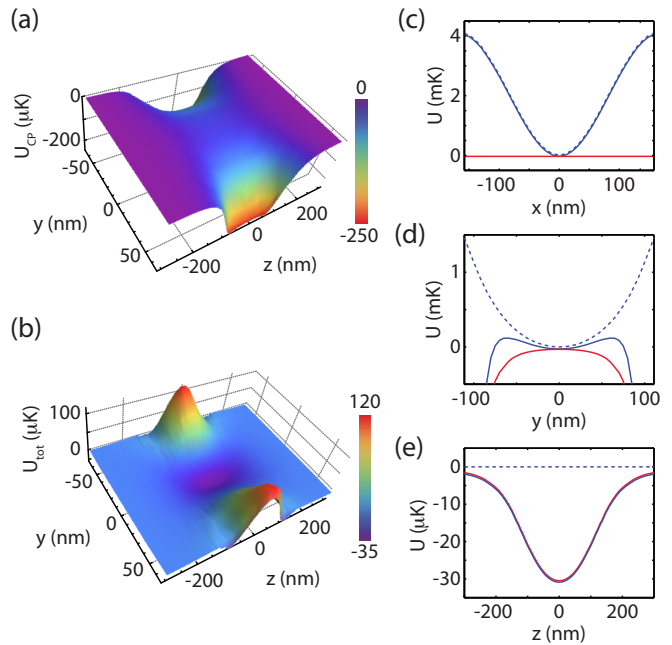}
\vspace{-5mm}
\caption{Trapping potentials for the double nanobeam structure in Fig. \ref{fig1}(e) for Cs $6S_{1/2}, F = 4$ level and $\lambda_{T} = 793$nm. (a) Casimir-Polder potential $U_{\mathrm{CP}}(\mathbf{r})$ and (b) total potential $U_{\mathrm{tot}}(\mathbf{r})=U_{\mathrm{CP}}(\mathbf{r})+U_{\mathrm{FORT}}(\mathbf{r})$ in the transverse $x=0$ plane. (c-e) show line cuts of $U_{\mathrm{CP}}$ (red solid), $U_{\mathrm{FORT}}$ (blue dashed), and $U_{\mathrm{tot}}$ (blue solid) along the (c) $x$-, (d) $y$-, and (e) $z$-axis. Average trap intensity for a unit cell is $3.5$mW/$\mu$m$^2$.}
\vspace{-5mm}
\label{fig3}%
\end{figure}

Figures \ref{fig2} and \ref{fig3} display numerical results for $U_{\mathrm{CP}}(\mathbf{r})$, $U_{\mathrm{FORT}}(\mathbf{r})$, and $U_{\mathrm{tot}}(\mathbf{r})$ for the single and double nanobeams for $k_x$ below the $X$-point. The calculations are for the $6S_{1/2}, F = 4$ hyperfine ground state of Cs for the FORT modes indicated in Fig. \ref{fig1}(c, g) \cite{vector_shifts}. For these initial calculations, we make the reasonable assumption for $SiN$ that the dielectric constant $\epsilon$ is frequency independent, $\epsilon(\mathbf{r},\omega) \rightarrow \epsilon(\mathbf{r})$.

$U_{\mathrm{tot}}(\mathbf{r})$ for the single nanobeam in Fig. \ref{fig2} reveals that modest optical intensity is sufficient to overcome the attractive CP interactions and create a stable potential minimum in the center of the vacuum space at $\mathbf{r}_{min}=0$ within a unit cell. An atom would be localized at distances $(d_x,d_y)=(123,373)$nm from the walls of the dielectric. The trap oscillation frequencies for a Cs atom would be $(f_x,f_y,f_z)\simeq(612,180,484)$kHz.

For the double nanobeams \cite{Eichenfeld:2009}, the FORT alone is insufficient to trap the atom, as the mode has a (weak) local intensity maximum along the $z$ direction that repels an atom. However, the CP potential $U_{\mathrm{CP}}(\mathbf{r})$ along $z$ provides the force necessary to overcome the repulsive optical force and to form a stable trap. The result is a hybrid optical-vacuum trap that circumvents the `no-go' theorem for vacuum trapping alone \cite{Kardar}.

Potentials for our hybrid trap are illustrated in Fig. \ref{fig3}. At the trap minimum $\mathbf{r}_{min}=0$, an atom would be localized at distance $d_y=125$nm from adjacent surfaces of the dielectric beams. Oscillation frequencies for a Cs atom would be $(f_x,f_y,f_z)\simeq(1013,390,57)$kHz.

As concerns strong radiative interactions, our structures trap an atom in a region of large amplitude for the probe field, leading to small mode volume per unit cell \cite{SM}. It is well known that atom-photon interactions can be further enhanced near a band edge \cite{Soljacic2002,Koenderink2006,Rao2007,Baba2008,Hoang2012}, where the density of states diverges due to a van Hove singularity. To quantify the radiative coupling, we determine the decay rate $\gamma_{\mathrm{tot}}$ for a point dipole located at $\mathbf{r}_a=0$ for a structure with $N$ unit cells \cite{finitecell}. FDTD calculations are performed to evaluate the classical Green's tensor $\mathbf{G}(\mathbf{r}_a,\mathbf{r}_a,\omega)$ and thence $\gamma_{\mathrm{tot}}$ following Refs. \cite{Agarwal1975, Buhmann2004, Sondergaard04}, as described in \cite{SM}.

Figures~4(a) and 5(a) display the diagonal components $Im[G_{ii}(\nu_d)]$ of the Green's tensor as functions of dipole frequency $\nu_d=\omega_d/2\pi$ and relate to the emission rate of resonant point dipoles polarized along the $i=x$, $y$, or $z$-axis for the single and double nanobeams.
Firstly, in Fig.~4(a), $Im[G_{xx}]$ is enhanced along the $x$-(periodic) direction across a broad frequency range, and is suppressed in the $y$- and $z$- directions, as can be explained by the orientation of the induced array of image dipoles along the single nanobeam. When the source dipole is polarized along the $x$-axis, the image dipoles line up head-to-tail and, just below the $X$-point, constructively interfere to enhance dipole emission; $y,z$ orientations render destructive interference and suppressed emission. Comparable suppression is not apparent for the double nanobeams in Fig.~5(a) since no such array of image dipoles is formed.

Secondly, Figs. 4(a) and 5(a) display a series of resonant peaks due to strong emission into various guided modes. In the region near the Cs $D2$ line (i.e., $\nu_d \simeq \nu_A=\omega_A/2\pi = 352$~THz), we find peaks in $Im[G_{xx}]$ for the single nanobeam and in $Im[G_{yy}]$ for the double nanobeam. These peaks are due to emission into our designated probe modes for the respective structures, where for the single (double) nanobeam(s), the probe mode is principally polarized along the $x$-($y$-) axis. Each peak is from a discrete set of propagation constants $k_x^{(n)} \simeq \pi n/aN$ imposed by the boundary conditions for the finite structures. Here, $N$ is the total number of cells in the single (double) beam, and $n \leq N$ is an even (odd) integer. We find excellent agreement between the frequencies of these resonances and the band diagram of the probe mode (`diamonds' in Figs.~1(b) and (f)) for various values of $(n,N)$.

The peaks become larger and narrower as $k_x^{(n)}$ approaches the $X$-point, owing to the diminishing group velocity \cite{SM,Rao2007,Hoang2012}. Beyond the $X$-point, the probe resonances disappear, leaving a broad background corresponding to coupling into lossy (radiation) modes.

\begin{figure}[t!]
\centering
\includegraphics[width=1.0\columnwidth]{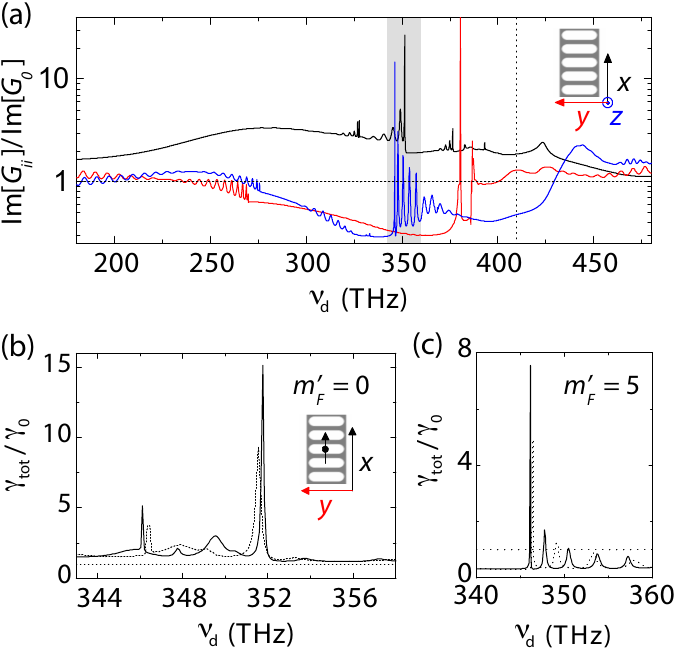}
\vspace{-3mm}
\caption{Green's tensor and total atomic decay rate $\gamma_{\mathrm{tot}}$ versus source dipole frequency $\nu_d$ for the single nanobeam at $\mathbf{r}=0$. (a) shows the diagonal components of the Green's tensor, $Im[G_{xx}]$ (solid black), $Im[G_{yy}]$ (red), and $Im[G_{zz}]$ (blue), normalized to the free space value $Im[G_0]$ (dashed line). The number of unit cells is $N=81$. Cesium $D2$-line frequency $\nu_A=352$~THz is centered in the shaded area. The vertical dotted line marks the light line, beyond which all decay channels are lossy. (b, c) show $\gamma_\mathrm{tot}$ (black curves), normalized to the free space value $\gamma_0$ (dotted line), in the frequency range marked by the shaded area in (a). The solid (dashed) curve is evaluated using $81$($61$) unit cells. The atomic spin is aligned to the $x$-axis, with the spin projection quantum number (b) $m_F^{\prime}=0$ and (c) $m_F^{\prime}=5$.}
\vspace{-6mm}
\label{fig4}%
\end{figure}

On an expanded frequency scale around $\nu_d \simeq \nu_A$, Figs.~4(b) and 5(b) show calculated atomic decay rates $\gamma_{\mathrm{tot}}$ for the $6P_{3/2},F^{\prime}=5 \rightarrow 6S_{1/2},F=4$ transition in atomic Cs \cite{SM}. When the atomic dipole is aligned along the principal polarization of the designated probe mode ($\hat{x}$ for the single beam and $\hat{y}$ for the double beam), the emission rate $\gamma_{1D}$ into the probe mode is strongly enhanced at frequencies corresponding to $k_x^{(n)}$ near the $X$-point. Specifically, for $\nu_d = \nu_A$ large enhancements in $\gamma_{1D}$ occur for the initial excited state $6P_{3/2},F^{\prime}=5,m_F^{\prime}=0$, while $\gamma_{1D}$ is suppressed for the initial state $m_F^{\prime}=5$. This is because the probe mode predominantly supports $\pi$-polarization and hence $\Delta m_F=0$. Coupling between states with $\Delta m_F\neq0$ is small. Of course, additional guided modes can contribute to $\gamma_{\mathrm{tot}}$, as is evidenced for the $m_F^{\prime}=5$ state due to field polarizations perpendicular to the atomic spin, such as $\hat{z}$ ($\hat{x}$) for the single (double) beam(s) in Fig.~4(a) (5(a)).

\begin{figure}[t!]
\centering
\includegraphics[width=1.0\columnwidth]{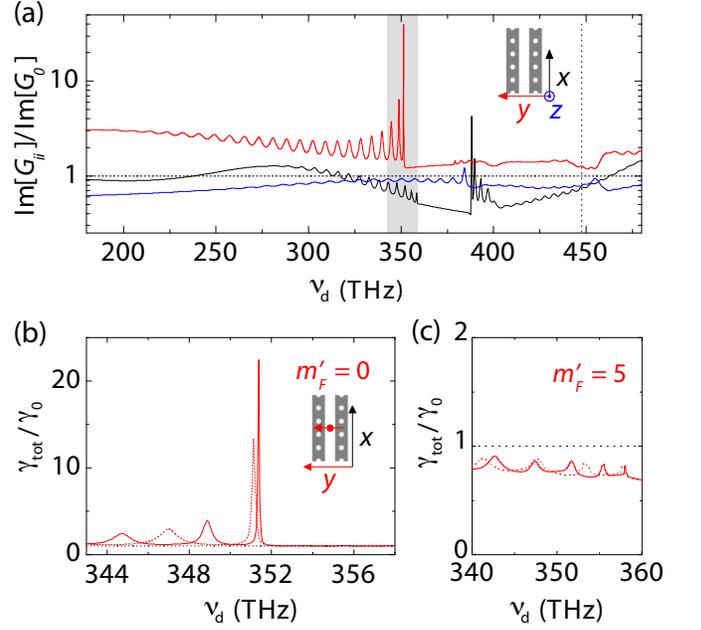}
\vspace{-3mm}
\caption{As in Fig. \ref{fig4} now for the double nanobeam at $\mathbf{r}=0$. (a) shows the diagonal components of the Green's tensor, $Im[G_{xx}]$ (solid black), $Im[G_{yy}]$ (red), and $Im[G_{zz}]$ (blue), normalized to $Im[G_0]$ for free space (dashed line). $N=81$ unit cells. Cesium $D2$-line frequency $\nu_A=352$~THz is centered in the shaded area. The vertical dotted line marks the light line, beyond which all decay channels are lossy. (b, c) show $\gamma_\mathrm{tot}$ (red curves), normalized to the free space value $\gamma_0$ (dotted line), in the frequency range marked by the shaded area in (a). The solid (dashed) curve is evaluated using $81$($61$) unit cells. The atomic spin is aligned to the $y$-axis, with the spin projection quantum number (b) $m_F^{\prime}=0$ and (c) $m_F^{\prime}=5$.}
\label{fig5}%
\vspace{-5mm}
\end{figure}

From $ \gamma_\mathrm{tot}(\nu_d)$ and an analytic model of coupling to the guided-mode near the $X$-point, we estimate the contributions of $\gamma_{1D}$ and $\gamma'$ to $\gamma_{\mathrm{tot}}=\gamma_{1D}+\gamma^{\prime}$ near the largest resonances in Figs. \ref{fig4}(b) and \ref{fig5}(b) \cite{SM}. For $m_F^{\prime}=0$ and $N=81$, we find that $\gamma_{1D}/\gamma_0 \simeq 15$ and  $\gamma^{\prime}/\gamma_0 \simeq 1.2$ for the single nanobeam, while $\gamma_{1D}/\gamma_0 \simeq 21$ and $\gamma^{\prime}/\gamma_0 \simeq 1.0$ for the double nanobeams \cite{compare}. Here, $\gamma_0/2\pi=5.2$MHz, the free-space Cs decay rate.

The ratios $\gamma_{1D}/\gamma_{tot}$ and $\gamma_{1D}/\gamma^{\prime}$ serve as metrics for the strength of atom-photon interactions for our $1D$ photonic crystals. For example, the resonant reflectivity $r_0$ of a trapped atom for the probe field should scale as $r_0 = \gamma_{1D}/\gamma_{tot}$ \cite{fan05,Chang2012}, which for the double nanobeams leads to $r_0 \simeq 0.95$. For a cavity QED system with one `impurity' atom surrounded by $N_A$ `mirror' atoms along a $1D$-lattice \cite{Chang2012}, the ratio of the coherent coupling rate $g_1 = \sqrt{N_A} \gamma_{1D} /2$ to the effective dissipative rate $\gamma^{\prime}$ would exceed unity even for $N_A=1$ atom.
For conventional cavity QED, we estimate a vacuum Rabi frequency $\sim 2\pi\times6$GHz for an atom trapped in the $1D$-photon
crystal waveguides studied here, making the reasonable assumption of a cavity formed from $N\sim10$ unit cells.


Certainly there are challenges to the implementation of our designs for trapping atoms in $1D$ photonic crystal waveguides with strong single-photon interactions, including atom loading into the small trap volume and light scattering from device imperfections. We are working to address these issues by numerical simulation, device fabrication, and cold-atom experiments with nanoscopic structures. Our efforts are motivated by the prediction $\gamma_{1D}/\gamma_{tot} \gtrsim 0.9$ in Figs. \ref{fig4}, \ref{fig5}, which is unprecedented in AMO physics and which could create new scientific opportunities (e.g., quantum many-body physics for $1D$ atom chains with photon-mediated interactions, and high-precision studies of vacuum forces). Moreover, our double nanobeam structure provides proof-of-principle for a promising new concept that combines optical and vacuum forces to form stable traps for neutral atoms in dielectric nanostructures.

We gratefully acknowledge the contributions of D. J. Alton, K. S. Choi, D. Ding, and A. Goban.
Funding is provided by the IQIM, an NSF Physics Frontier Center with support of the Moore Foundation, by the AFOSR QuMPASS MURI, by the DoD NSSEFF program (HJK), and by NSF Grant PHY0652914 (HJK). DEC acknowledges funding from Fundacio Privada Cellex Barcelona.

$^\ast$ These authors contributed equally to this research.

\newpage

\section{Supplemental Material}
\vspace{5mm}

\section{Calculation of Casimir-Polder Potentials}
\vspace{-2mm}

The Casimir-Polder potential $U_{\mathrm{CP}}(\mathbf{r}_a)$ in Ref. \cite{manuscript} is calculated from the following integral \cite{SMBuhmann2004}:
\begin{multline}
U_{\mathrm{CP}}(\mathbf{r}_a) =\\
- \frac{\hbar \mu_0}{2 \pi} \mathrm{Im} \left\{  \int_0^\infty d \omega \omega^2 \mathrm{Tr} [ \bm{\alpha}^0(\omega) \cdot  \mathbf{G}_{\mathrm{sc}}(\mathbf{r}_a,\mathbf{r}_a,\omega)] \right\},\label{ucp}
\end{multline}
where $\mathrm{Tr}[.]$ denotes the trace, $\bm{\alpha}^0$ is the dynamic polarizability tensor of ground-state Cesium atom, and $\mathbf{G}_{sc}(\mathbf{r}_a,\mathbf{r}_a,\omega) = \mathbf{G}(\mathbf{r}_a,\mathbf{r}_a,\omega) - \mathbf{G}_0(\mathbf{r}_a,\mathbf{r}_a,\omega)$ is the scattering Green's tensor, that is, the Green's tensor $\mathbf{G}$ subtracted by the vacuum contribution $\mathbf{G}_0$ evaluated at atomic location $\mathbf{r}_a$; $2\pi\hbar$ is Planck's constant, and $\mu_0$ is the vacuum permeability. The Green's tensor is the solution to the Maxwell equation $\Big[ \nabla \times \nabla \times - \frac{\omega^2}{c^2} \epsilon (\mathbf{r},\omega) \Big] \mathbf{G} (\mathbf{r} ,\mathbf{r} ',\omega) = \mathbf{I} \delta^{(3)} (\mathbf{r} -\mathbf{r} ')$, corresponding to the electric field response to a point dipole current source. $\epsilon(\mathbf{r}, \omega)$ is the dielectric function, and $\mathbf{I}$ is the unity tensor.

We employ finite-difference-time-domain (FDTD) calculations \cite{SMMEEP} to solve numerically for the Green's tensors of our structures. The integral of Eq.~(\ref{ucp}) is evaluated by adapting a procedure established in Ref. \cite{SMRodriguez2009} and by using a deformed contour $\omega (\xi) = \xi \sqrt{1+i\sigma/\xi}$ in the upper half of the complex frequency plane, parametrized by a real number $ \xi \ge 0$ and a constant $\sigma > 0$. As explained in Ref. \cite{SMRodriguez2009}, this is equivalent to solving the Green's tensor at real frequencies $\xi$ with a fictitious global conductivity applied to the dielectric function $\epsilon'(\mathbf{r},\xi) = (1+ i \sigma /\xi) \epsilon(\mathbf{r}, \xi)$. The integration can then be performed in the time domain (via the convolution theorem) and converges quickly due to fast decay from $\sigma$.

Specifically, Eq.~(\ref{ucp}) is numerically evaluated using
\begin{equation}
U_{\mathrm{CP}} (\mathbf{r}_a) = \frac{\hbar}{2 \pi}  \int_{0}^{\infty} dt  \mathrm{Im} \left[g_{\mu\nu}(-t)\right] \hat{x}_\mu \cdot \mathbf{E}_{\mathrm{sc},\nu}(\mathbf{r}_a,t), 
\end{equation}
where $\mathbf{E}_{\mathrm{sc},\nu}(\mathbf{r}_a,t)$ is the (real) electric field generated by a point dipole current source $\mathcal{J}=\delta(t) \hat{x}_\nu$ ($\hat{x}_\nu = \hat{x}, \hat{y}, \hat{z}$) located at the position $\mathbf{r}_a$ and scattered by a structure with a dielectric function $\epsilon'(\mathbf{r}, \xi)$  \cite{SMperiodic}. Here, the indices $\mu$ and $\nu$ are repeated for summation convention, $g_{\mu\nu}(t)$ is the Fourier transform of $g_{\mu\nu}(\xi) = -i \xi \sqrt{1+\frac{i \sigma}{\xi}} \left( 1+\frac{i \sigma}{2 \xi} \right) \Theta(\xi)  \bm{\alpha}^0_{\mu\nu}(\omega(\xi))$,  and $\Theta(\xi)$ is the Heaviside step function. For the initial calculations in Ref.~\cite{manuscript}, we take the dielectric constant $\epsilon$ to be frequency independent, $\epsilon (\mathbf{r},\omega) \rightarrow \epsilon(\mathbf{r})$.

\section{Calculation of $\gamma_{tot}$}

To determine the total spontaneous decay rate $\gamma_{\mathrm{tot}}$ for an atom in our structures, we also solve for the classical Green's tensors and evaluate $\gamma_{\mathrm{tot}}$ via \cite{SMBuhmann2004,SMAgarwal1975,SMSondergaard04}
\begin{eqnarray}
\gamma_\mathrm{tot} = \frac{2\mu_0 \omega_j^2}{\hbar} \mathrm{Im} \left\{ \sum_{\{0\}} \mathrm{Tr} [ \mathbf{D}_j \cdot \mathbf{G}(\mathbf{r}_a,\mathbf{r}_a,\omega_j) ] \right\},\label{decay}
\end{eqnarray} 
where $\mathbf{D}_j = \langle \{0\} | \mathbf{d}^\dagger | j \rangle \langle j | \mathbf{d} | \{0\} \rangle$ is the dipole matrix element between the ground state manifold and the excited state $j$, and $\omega_j$ is the transition frequency. The total decay rate $\gamma_{\mathrm{tot}} = \gamma_\mathrm{1D} + \gamma'$ includes the decay rate $\gamma_\mathrm{1D}$ to a guided mode of interest as well as the rate  $\gamma'$ to all other modes of the structure, including lossy modes. As discussed below, the contributions of $\gamma_\mathrm{1D} ,\gamma'$ to $\gamma_{\mathrm{tot}}$ can be estimated from the global frequency dependence $ \gamma_\mathrm{tot}(\omega)$.

To obtain $\gamma_{tot}(\omega)$, we evaluate the Green's tensor for the real dielectric function $\epsilon(\mathbf{r})$ using the FDTD method, followed by a discrete Fourier analysis. 

\vspace{-2mm}
\section{Validation}
 
To validate our numerical procedures, we have performed calculations of $U_{\mathrm{CP}}$ for several geometries where analytical solutions are available, including an atom near an infinite dielectric or metal half-space \cite{SMDzyaloshinskii1961} and an atom located above an infinite dielectric grating \cite{SMContrerasReyes2010}, and found excellent agreement between our simulations and the exact results.

We have validated our calculations of $\gamma_{\mathrm{tot}}$ for the cases of an (atomic) dipole near an infinite dielectric, metallic parallel plates, a nanofiber \cite{SMKien05}, and $2D$-photonic band-gap microcavities \cite{SMHwang99}.

\vspace{-2mm}
\section{Guided Mode Resonances}
 
For our structure with an infinite number of unit cells, a guided mode (denoted by $\lambda$) contribution to the imaginary part of the Green's tensor can be calculated as \cite{SMSondergaard04, SMgmode}, $\mathrm{Im} \left[\mathbf{G}^{\lambda}_\mathrm{1D}(\mathbf{r}_a,\mathbf{r}_a,\omega)\right] = ac^2 \mathbf{u}_{\lambda}(\mathbf{r}_a;k_x) \otimes \mathbf{u}^*_{\lambda}(\mathbf{r}_a;k_x)/2\omega v_\lambda$, when the frequency $\omega$ intersects the frequency band $\omega_\lambda$ at a propagation constant $k_x$ below the light line. Here, $\mathbf{u}_{\lambda}(\mathbf{r};k_x)$ is the orthonormal mode function, and $v_\lambda$ is the group velocity, both available via numerical calculations \cite{SMMPB, SMCOMSOL}. As we scan the frequency $\omega$, $\mathrm{Im} \left[\mathbf{G}^{\lambda}_\mathrm{1D}\right]$ increases monotonically and diverges as $k_x$ approaches the $X$-point, where $v_\lambda \rightarrow 0$. The guided mode Green's tensor vanishes when $\omega$ lies beyond the frequency of the band edge. 

Based on this analysis, we can evaluate the decay rate $\gamma_{\mathrm{1D}}^{(\infty)}(\omega)$ into the designated probe mode for an infinite structure, and compare it with the heights of resonant features in $\gamma_\mathrm{tot}(\omega)$ for finite structures with different numbers of unit cells, as shown in Figs.~4(b) and 5(b) of Ref. \cite{manuscript}. Indeed, the actual $\gamma_\mathrm{1D}$ of a finite-size structure must deviate from $\gamma_{\mathrm{1D}}^{(\infty)}$ due to boundary conditions that transform a continuous spectrum into a discrete set of resonant peaks \cite{SMRao07}, as shown in Figs.~4(b) and 5(b) \cite{manuscript}.  When the number of unit cells is increased in our calculation of $\gamma_{\mathrm{1D}}$ over the range $N=11$ to $N=81$, we find that the frequencies $\omega^{(n)}$ of the resonant peaks shift in position and the peaks change height. As documented by the black diamonds in Figs. 1(b), 1(f) of Ref. \cite{manuscript}, the $\omega^{(n)}$ arise from the discrete set of propagation constants $k_x^{(n)} \simeq \pi n/aN$ imposed by the boundary conditions for the finite structures with $n$ either even or odd.

The peaks in $\gamma_\mathrm{tot}$ at the set of frequencies $\omega^{(n)}$ build up on top of a fairly constant background within the frequency range displayed in Figs.~4(b) and 5(b) of Ref. \cite{manuscript}. We assume that this background represents the contribution of $\gamma^{\prime}$ to $\gamma_\mathrm{tot}$, and subtract the background to estimate $\gamma_{\mathrm{1D}}$. The resulting form for $\gamma_{\mathrm{1D}}(\omega)$ consists of a set of resonant peaks whose heights at discrete $\omega^{(n)}$ qualitatively map out $\gamma_{\mathrm{1D}}^{(\infty)}$ calculated for the infinite structure, with the maximum peak height for $\gamma_{\mathrm{1D}}$ occurring for the peak closest to the band edge. Moving further away from the band edge, we find that our numerical estimate of $\gamma_\mathrm{tot}(\omega)-\gamma^{\prime}(\omega)$ asymptotes to the calculated value of $\gamma_{\mathrm{1D}}^{(\infty)}(\omega)$ reasonably well.

\vspace{-3mm}
\subsection{Estimation of $\gamma_\mathrm{1D}$ and $\gamma^{\prime}$}
\vspace{-2mm}

From the previous discussions, we identify that the decay rate into other modes $\gamma^{\prime}$ can be read off from the broad background in $\gamma_\mathrm{tot}$. Specifically, we estimate $\gamma' = \gamma_\mathrm{tot} (\omega')$ at a frequency $\omega'$ just across the band edge and away from any resonant peak for a guided mode. The decay rate into the probe mode $\gamma_\mathrm{1D}$ can then be estimated using $\gamma_\mathrm{1D}=\gamma_\mathrm{tot}-\gamma'$.

For the single-beam structure and the atomic spin orientation shown in Fig.~4(b)~\cite{manuscript}, we find a peak total decay rate $\gamma_\mathrm{tot}/\gamma_0 \approx 15$ and a background level $\gamma'/\gamma_0 \approx  1.2$ near the Cesium $D2$-line frequency $\nu_A = 352~$THz. We estimate the coupling to the resonant probe mode $\gamma_\mathrm{1D} = \gamma_\mathrm{tot} - \gamma' \approx 14\gamma_0$. For the double-beam structure and spin orientation shown in Fig.~5(b), we find $\gamma_\mathrm{tot}/\gamma_0 \approx 22$, $\gamma'/\gamma_0 \approx  1$, and, therefore, $\gamma_\mathrm{1D} /\gamma_0 \approx21$. Here, $\gamma_0/2\pi=5.2$MHz is the free-space (vacuum) decay rate for the $D2$ line.

\vspace{-1mm}
\section{Effective Area and Mode Volume for Probe}
\vspace{-2mm}

Both the single and double nanobeam structures in Ref. \cite{manuscript} lead to atom localization in a region of large amplitude for the probe field. One measure of the strength of the atom-field coupling is the effective mode volume $V_m$ per unit cell, where
\begin{equation}
V_m=\int \epsilon(\mathbf{r}) |\mathbf{E}(\mathbf{r})|^2 d^3r /\epsilon(\mathbf{r}_{min}) |\mathbf{E}(\mathbf{r}_{min})|^2.
\label{Vm}
\end{equation}
Here the integration is carried out over the volume of a unit cell. That is, the integration domain along propagation direction $x$ extends over the distance $a$ (i.e., the lattice constant), while in the transverse $y,z$ directions, the integration domain is from $-\infty$ to $+\infty$.

For the single nanobeam, the probe mode has a global maximum at $\mathbf{r}_{min}=0$ and an effective mode volume $V_m\sim 0.13~\mu$m$^3$. For the double nanobeams, the probe mode has a saddle-like intensity distribution around $\mathbf{r}_{min}=0$, resulting in $V_m \sim 0.11~\mu$m$^3$ for a unit cell.

\vspace{3mm}
$^\ast$These authors contributed equally to this work.



\begin{thebibliography}{10}
\bibitem{redFORT} We have also designed structures for trap and probe modes for red-detuned FORTs with an atom trapped at the intensity maximum in a unit cell.
\bibitem{vector_shifts} $U_{\mathrm{tot}}(\mathbf{r})$ is largely insensitive to the particular atomic Zeeman sublevel owing to negligible vector shifts due to the high degree of linear polarization of the trapping modes and to the absence of ground state tensor polarizability.
\bibitem{finitecell} A finite number of unit cells is chosen to better represent an actual structure. Since our focus is not on specific boundaries, the structure is terminated by vacuum.
\end{thebibliography}
\begin{thebibliography}{10}
\bibitem{SMperiodic} To obtain the CP potential for a single atom inside a periodic structure, we impose Bloch-periodic boundary conditions on a single unit-cell in the FDTD calculations. We then sum over all fields $\mathbf{E}_{\mathrm{sc},\nu} (\mathbf{r}_a, t;k_x)$ with the Bloch wavevector $k_x$ across the first Brillouin zone to obtain $\mathbf{E}_{\mathrm{sc},\nu} (\mathbf{r}_a,t) = \frac{a}{2\pi} \int_{-\pi/a}^{\pi/a} dk_x \mathbf{E}_{\mathrm{sc},\nu} (\mathbf{r}_a, t;k_x)$, which is mathematically equivalent to solving for the fields of a single point dipole in an infinite periodic structure. The integral is approximated using a 10 point Gaussian quadrature.
\end{thebibliography}
\end{document}